\documentclass[12pt]{article}
\usepackage{amsmath,amssymb}

\newcommand{\be}{\begin{equation}}
\newcommand{\ee}{\end{equation}}
\newcommand{\ea}{\end{array}}
\newcommand{\beqa}{\begin{eqnarray}}
\newcommand{\eeqa}{\end{eqnarray}}

\newcommand{\nn}{\nonumber}
\def\tr{\mathop{\rm Tr}\nolimits}
\def\BI{{\rm 1\!l}}
\def\half{\frac{1}{2}}

\newcommand{\gapproxeq}{\lower .7ex\hbox{$\;\stackrel{\textstyle
>}{\sim}\;$}}
\newcommand{\lapproxeq}{\lower .7ex\hbox{$\;\stackrel{\textstyle
<}{\sim}\;$}}

\newcounter{appendice}

\def\thebibliography#1{{\bf REFERENCES\markboth
 {REFERENCES}{REFERENCES}}\list
 {[\arabic{enumi}]}{\settowidth\labelwidth{[#1]}\leftmargin\labelwidth
 \advance\leftmargin\labelsep
 \usecounter{enumi}}
 \def\newblock{\hskip .11em plus .33em minus -.07em}
 \sloppy
 \sfcode`\.=1000\relax}

\begin{document}
\begin{titlepage}
\title{{\small\hfill SU-4252-769,  DFUP-02-14 }\\ \medskip
The Fuzzy Ginsparg-Wilson Algebra: A Solution of the Fermion Doubling
Problem  }
\author{
A.P. Balachandran$^a$, Giorgio Immirzi$^b$, \\
{\small\it $^a$ Physics Department,
Syracuse University}\\
{\small\it Syracuse NY 13244-1130, USA}\\
{\small\it $^b$ Dipartimento di Fisica, Universit\`a di Perugia 
{\small\rm and} INFN, Sezione di Perugia,}\\
{\small\it Perugia, Italy}\\
}
\maketitle
\begin{abstract}
The Ginsparg-Wilson algebra is the algebra underlying the Ginsparg-Wilson
solution of the fermion doubling problem in lattice gauge theory. The Dirac 
operator of the fuzzy sphere is not afflicted with this problem. Previously
we have indicated that there is a Ginsparg-Wilson operator underlying
it as well in the absence of gauge fields and instantons. Here we
develop this observation systematically and establish a Dirac operator
theory for the fuzzy sphere with or without gauge fields, and always
with the Ginsparg-Wilson algebra. There is no fermion doubling in this
theory. The association of the Ginsparg-Wilson algebra with the fuzzy
sphere is surprising as the latter is not designed with this algebra
in mind. The theory reproduces the integrated $U(1)_A$ anomaly and
index theory correctly.
\end{abstract}
\end{titlepage}

\section{Introduction}

A central task in lattice approximation to quantum field theories
(qft's) is the treatment of chiral fermions. General theorems due to
Nielsen and Ninomiya and others \cite{nielsen} reveal a serious
obstruction to their rigorous formulation on a lattice. As the
standard model involves chiral fermions, there is thus a fundamental
difficulty with lattice approximations.

Years ago,  Ginsparg and  Wilson \cite{ginsparg} proposed an
approximate manner to overcome this difficulty. In the original
formulation, it is based on a Dirac and chirality operator fulfilling
particular algebraic relations. In the continuum limit, anticommuting
Dirac and chirality operators can be obtained therefrom. The
Ginsparg-Wilson method is an effective tool in the theoretical
analysis of lattice theories and reproduces important topological effects
like chiral anomalies in an approximate manner. 

Fuzzy physics \cite{madore}\cite{bal1} concerns an approach to
regulating qft's which can be an alternative to lattice methods. It
gives finite-dimensional matrix approximations to qft's and
incorporates ideas of non-commutative geometry \cite{connes}. It has a
well-articulated theory of Dirac operator for the fuzzy sphere which
approximates the continuum Dirac operator very well and also
reproduces the correct index theory and chiral anomaly. Subtle
topological features like instantons and complex structures can be
formulated \cite{baez}. Chiral fermions too can be described with no
fermion doubling \cite{bal2}. For fuzzy ${\mathbb
C}P^N$-models as well, the Dirac operator to the extent investigated
\cite{grosse} seems an excellent approximation to the continuum Dirac
operator and capable of reproducing significant topological features
of the continuum.

In a previous paper \cite{bal2}, we reported our joint work on the
Dirac operator of \cite{klimcik} for the fuzzy sphere. 
Here we establish that the `free'
fuzzy Dirac operator in the absence of instantons fulfills the
defining relations of the Ginsparg-Wilson algebraic system. This
result has a strong element of surprise as fuzzy physics is not
consciously designed to fulfill such relations. 

In this paper we review our previous work and extend it to cover gauge
fields and instanton sectors. This extension has a new formulation of
the Dirac operator on the fuzzy sphere, and is based on an appropriate
realization of the Ginsparg-Wilson algebra. This Dirac operator has
several positive features. Its spectrum in the absence of gauge field
fluctuations is precisely that in the continuum below
 a suitable angular momentum cut-off. There is no correction
whatever to spectrum below the cut-off. There is no fermion doubling
and chiral fermions can be effortlessly treated. The $U(1)_A$ anomaly
in the integrated form is reproduced exactly. We have not looked at
its local form, but its treatment in alternative approaches exists
\cite{peter}\cite{bg}.

For other work applying the Ginsparg-Wilson approach to the fuzzy 
sphere, see \cite{aoki}.

While these are points in favour of our approach, it appears that
the Ginsparg-Wilson approach, either in the lattice or fuzzy physics
context, is not easy to adapt to numerical work. This is a serious
difficulty and has to be overcome.

\section{A Review of the Ginsparg-Wilson Algebra.}
\setcounter{equation}{0}

We follow \cite{bal1}\cite{bal2} in this presentation.

In its generality, the Ginsparg-Wilson algebra $\cal A$ can be
defined as the unital $*$-algebra over $\mathbb C$ generated by two
$*$-invariant involutions $\Gamma$ and $\Gamma'$:
\be
{\cal A}=\langle\Gamma,\Gamma':\quad \Gamma^2={\Gamma'}^2=\BI,\quad 
\Gamma^*=\Gamma,\quad {\Gamma'}^*=\Gamma'\rangle\ ,
\label{gwfi}
\ee
$*$ denoting the adjoint.
The unity of $\cal A$ has been indicated by $\BI$.

In any such algebra, we can define a Dirac operator
\be 
D'=\frac{1}{a}\Gamma(\Gamma+\Gamma')\ ,
\label{gwfii}
\ee
where $a$ is the lattice spacing. It fulfills
\be
{D'}^*=\Gamma\, D'\,\Gamma,\quad [\Gamma,D']_+=a\,D'\,\Gamma\,D'\ .
\label{gwfiii}
\ee
(\ref{gwfii}) and (\ref{gwfiii}) give the original formulation
\cite{ginsparg}.  But they are equivalent to (\ref{gwfi}), since 
(\ref{gwfii}) and (\ref{gwfiii}) imply that
\be
\Gamma'=\Gamma(aD')-\Gamma
\label{gwfiv}
\ee
is a $*$-invariant involution \cite{luscher}\cite{fujikawa}.

Each representation of (\ref{gwfi}) is a particular realization of the
Ginsparg-Wilson algebra. Representations of physical interest are
reducible.

In our work we choose
\be
D=\frac{1}{a}(\Gamma+\Gamma')\ ,
\label{gwfv}
\ee
instead of $D'$ as our Dirac operator, as it is self-adjoint and has
the desired continuum limit.

From $\Gamma$ and $\Gamma'$, we can construct the following elements of
$\cal A$:
\beqa
\Gamma_0&=&\frac{1}{2}[\Gamma,\Gamma']_+\ ,   \label{gwfvi}\\
\Gamma_1&=&\frac{1}{2}(\Gamma+\Gamma')\ ,   \label{gwfvii}\\
\Gamma_2&=&\frac{1}{2}(\Gamma-\Gamma')\ ,   \label{gwfviii}\\
\Gamma_3&=&\frac{1}{2i}[\Gamma,\Gamma']\ .   \label{gwfix}
\eeqa

Let us first look at the centre ${\cal C}({\cal A})$ of $\cal A$ in
terms of these operators. It is generated by $\Gamma_0$ which commutes
with $\Gamma$ and $\Gamma'$ and hence with every element of $\cal A$\,.
$\Gamma_i^2,\ i=1,2,3$ also commute with every element of  $\cal A$,
but they are not independent of $\Gamma_0$. Rather,
\beqa
\Gamma_1^2=\frac{1}{2}(\BI+\Gamma_0) &,&\label{gwfx}\\
\Gamma_2^2=\frac{1}{2}(\BI-\Gamma_0) &,&\label{gwfxi}\\ 
\rightarrow\quad \Gamma_1^2+\Gamma_2^2=\BI\ &,&\label{gwfxii}\\
\Gamma_0^2+\Gamma_3^2=\BI\ &.&\label{gwfxiii}
\eeqa
Notice also that
\be
[\Gamma_i,\Gamma_j]_+=0\ ,\ i,j=1,2,3,\ i\ne j\ .
\label{gfwxiv}
\ee
From now on by $\cal A$ we will mean a representation of $\cal A$.

The relations (\ref{gwfx})-(\ref{gwfxiii}) contain spectral information.
From (\ref{gwfxiii}) we see that
\be
-1\le\Gamma_0\le 1\ ,
\label{gwfxvii}
\ee
where the inequality means that the eigenvalues of $\Gamma_0$ are
accordingly bounded. By (\ref{gwfx}), this implies that the eigenvalues
of $\Gamma_1$ are similarly bounded.

We now discuss three cases associated with (\ref{gwfxvii}).

\noindent{\bf Case 1} $\Gamma_0= \BI$.\ Call the subspace where 
$\Gamma_0= \BI$ as $V_{+1}$. On $V_{+1}$, $\Gamma_1^2=\BI$ and
$\Gamma_2=\Gamma_3=0$ by (\ref{gwfx}-\ref{gwfxiii}). This is subspace of
the top modes of the operator $|D|$.

\noindent{\bf Case 2} $\Gamma_0= -\BI$.\ Call the
subspace where $\Gamma_0= -\BI$ as $V_{-1}$. On $V_{-1}$, 
$\Gamma_2^2=\BI$ and
$\Gamma_1=\Gamma_3=0$ by (\ref{gwfx}-\ref{gwfxiii}). This is the
subspace of zero modes of the Dirac operator $D$.

\noindent{\bf Case 3} $\Gamma_0^2\ne \BI$. \ Call the subspace where 
$\Gamma_0^2\ne \BI$ as $V$. On this subspace,  $\Gamma_i^2\ne 0$ for
$i=1,2,3$ by (\ref{gwfix}-\ref{gwfxii}), and therefore  
\be
sign\,\Gamma_i=\frac{\Gamma_i}{|\Gamma_i|}\ ,\quad
|\Gamma_i|=\hbox{positive square root of}\ \Gamma_i^2
\label{gwfxv}
\ee
are well defined and by (\ref{gfwxiv}) generate a 
 Clifford algebra on $V$: 
\be
[sign\,\Gamma_i,sign\,\Gamma_j]_+=2\delta_{ij}\ .
\label{gwfxvi}
\ee

Consider $\Gamma_2$. It anticommutes with $\Gamma_1$ and $D$. Also
\be
\tr\,\Gamma_2=(\tr_V+\tr_{V_{+1}}+\tr_{V_{-1}})\Gamma_2\ ,
\label{gwfxviii}
\ee
where the subscripts refer to the subspaces over which the trace is taken.
These traces can be calculated:
\beqa
\tr_V\Gamma_2&=&\tr_V(sign\,\Gamma_i)\Gamma_2(sign\,\Gamma_i)
\quad (i\ \hbox{fixed,}\ \ne 2) \nn\\
&=&-\tr_V\Gamma_2\quad \hbox{by} (\ref{gwfxvi})\nn\\
&=&0,     \label{gwfxix}\\           
\tr_{V_{+1}}\Gamma_2&=&0,\quad \hbox{as}\ \Gamma_2=0\ \ \hbox{on}\ 
V_{+1}\ .
 \label{gwfxx}
\eeqa
So
\be
\tr\Gamma_2=\tr_{V_{-1}}\Gamma_2=\tr_{V_{-1}}(\frac{1+\Gamma_2}{2}-
\frac{1-\Gamma_2}{2})=\ \hbox{index of}\ \Gamma_1\ .
\label{gwfxxi}
\ee

Following Fujikawa \cite{fujikawa}, we can use $\Gamma_2$ as the
generator of chiral transformations. It is not involutive on 
$V\oplus V_{+1}$
\be 
\Gamma_2^2=\BI-\frac{\BI+\Gamma_0}{2}\ .
\label{gwfxxii}
\ee
But this is not a problem for fuzzy physics. In the fuzzy model below,
in the continuum limit, $\Gamma_0\to -\BI$ on all states with $|D|\le
$ a fixed `energy' $E_0$ independent of $a$ (and is $-\BI$ on $V_{-1}$
where $D=0$). We can see this as follows. $\Gamma_1=aD$, so that if
$|D|\le E_0, \ \Gamma_1\to 0$ as $a\to 0$. Hence by
(\ref{gwfx},\ref{gwfxii}), $\Gamma_0\to -\BI$ and $\Gamma_2^2\to\;\BI$
on these levels.

There are of course states, such as those of $V_{+1}$, on which
$\Gamma_2^2$ does not go to $\BI$ as $a\to 0$. But their (Euclidean)
energy diverges and their contribution to functional integrals
vanishes in the continuum limit.

We can interpret (\ref{gwfxxii}) as follows. The chiral charge of
levels with $D\ne 0$ gets renormalized in fuzzy physics. For levels
with $|D|\le E_0$, this renormalization vanishes in the naive
continuum limit.

We note that the last feature is positive: it resolves a problem in
previous work \cite{bal3}, where all the top modes had to be projected
out because of insistence that chirality squares to $\BI$ on $V_{+1}$;
see below.

For Dirac operators of maximum symmetry, $\Gamma_0$ is a function of
the conserved total angular momentum $\vec J$ as we shall show. It
increases with $\vec J^2$ so that $V_{+1}$ consists of states of
maximum $\vec J^2$. This maximum value diverges as $a\to 0$ as the
general argument above shows.

\section{ Fuzzy models}
\setcounter{equation}{0}
\subsection{ The Basic Algebra.}

The algebra for the fuzzy sphere characterized by cut-off $2L$ is the 
full matrix
algebra $Mat(2L+1)\equiv M_{2L+1}$ of $(2L+1)\times (2L+1)$ matrices. On
$M_{2L+1}$, the $SU(2)$ Lie algebra acts either on the left or on the
right. Call the operators for left action as $L^L_i$ and for right
action as $L^R_i$. We have
\be
L^L_ia=L_i a \ ,\  L^R_ia=aL_i\  ,\  a\in M_{2L+1}\ ,\nn
\ee
\be
 [L_i^L,L_j^L]=i\epsilon_{ijk}L^L_k\  , \quad
[L_i^R,L_j^R]=-i\epsilon_{ijk}L^R_k\ ,\quad 
(L_i^L)^2=(L_i^R)^2=L(L+1)\BI\ ,
\label{gwfxxiii}
\ee
where $L_i$ is the standard matrix for the $i$-th component of the
angular momentum in the the $(2L+1)$-dimensional irreducible
representation (IRR). The orbital angular momentum which becomes 
$-i(\vec r\wedge\vec\nabla)_i$ as $L\to\infty$ is
\be
{\cal L}_i=L_i^L-L_i^R\ ,\quad {\cal L}_ia=[L_i,a]\ .
\label{gwfxxiv}
\ee

As $L\to\infty$, both $\vec L^L/L$ and  $\vec L^R/L$ approach the unit
vector $\hat x$ with commuting components:
\be
\frac{\vec L^{L,R}}{L}\  
 {\lower .7ex\hbox{$\;\stackrel{\longrightarrow}
{\scriptstyle L\to\infty }\;$}}
\ \hat x\ ,\qquad \hat x\cdot\hat x=1\ ,\quad 
[\hat x_i,\hat x_j]=0\ .
\label{gwfxxiva}
\ee
$\hat x$ labels a point on the sphere $S^2$ in the continuum limit.

\subsection{The Fuzzy Dirac Operator (No Instantons or Gauge Fields)}

Consider $M_{2L+1}\otimes{\mathbb C}^2$. ${\mathbb C}^2$ is the
  carrier of the spin $1/2$ representation of $SU(2)$ with generators
  $\frac{1}{2}\sigma_i,\ \sigma_i=$ Pauli matrices. We can couple its
  spin $1/2$ and the angular momentum $L$ of $L^L_i$ to the value
  $L+1/2$. If $(1+\Gamma)/2$ is the corresponding projector, then
  \cite{bal2}\cite{watamura}
\be
\Gamma=\frac{\vec\sigma\cdot\vec L^L+1/2}{L+1/2} \ .
\label{gwfxxv}
\ee
$\Gamma$ is a self-adjoint involution,
\be
\Gamma^*=\Gamma\quad,\quad\Gamma^2=\BI\  .
\label{gwfxxvi}
\ee

There is likewise the projector $(\BI+\Gamma')/2$ coupling the spin
$1/2$ of ${\mathbb C}^2$ and the right angular momentum $-L^R_i$ to
$L+1/2$, where
\be
\Gamma'=\frac{-\vec\sigma\cdot\vec L^R+1/2}{L+1/2}={\Gamma'}^*\
,\quad{\Gamma'}^2=\BI\ . 
 \label{gwfxxvii}
\ee
The algebra $\cal A$ is generated by $\Gamma$ and $\Gamma'$.

The fuzzy Dirac operator of Grosse et al.\cite{klimcik} is
\be
D=\frac{1}{a}(\Gamma+\Gamma')=\frac{2}{a}\Gamma_1=\vec\sigma\cdot
(\vec L^L-\vec L^R)+1\ ,\quad
a=\frac{1}{L+1/2}\ .
\label{gwfxxviii}
\ee
Thus the Dirac operator is in this case  an element of the Ginsparg-Wilson 
algebra $\cal A$.

We can calculate $\Gamma_0$ in terms of $\vec J=\vec{\cal L}+\vec\sigma/2$:
\be
\Gamma_0=\frac{a^2}{2}[\vec J^2-2L(L+1)-\frac{1}{4}]\ .
\label{gwfxxix}
\ee
Thus the eigenvalues of $\Gamma_0$ increase monotonically with the
eigenvalues $j(j+1)$ of $\vec J^2$ starting with a minimum for $j=1/2$
and attaining a maximum of $1$ for $j=2L+1/2$.

$\Gamma_2$ is the chirality. It anticommutes with $D$. For fixed $j$, as
$L\to\infty$, $\Gamma_0\to -\BI$ and $\Gamma_2^2=\BI$ as expected. 
In fact, $\Gamma_2$ in the naive continuum limit  is the 
standard chirality for fixed $j$. As $L\to\infty, \ \Gamma_2\to
\sigma\cdot\hat x$.
As mentioned earlier, use of $\Gamma_2$ as chirality resolves a difficulty
addressed elsewhere \cite{bal2}\cite{bal3}, where $sign\,(\Gamma_2)$ was
used as chirality. That necessitates projecting out $V_{+1}$ and creates
a very inelegant situation.

Finally we note that there is a simple reconstruction of 
$\Gamma$ and $\Gamma'$ 
from their continuum limits \cite{ydri}. If $\vec x$ is not normalized, 
$\vec\sigma\cdot\hat x= \frac{\vec\sigma\cdot \vec x}{|\vec\sigma\cdot
 \vec x|},\ |\vec\sigma\cdot \vec x|\equiv|\big((\vec\sigma\cdot
\vec x)^2\big)^{1/2}|$.
As $\vec x$ can be represented by $\vec L^L$ or $\vec L^R$ in fuzzy physics,
natural choices for $\Gamma$ and $\Gamma'$ are $sign\,(\vec\sigma\cdot L^L)$
and $-sign\,(\vec\sigma\cdot L^R)$. The first operator is $+1$ on vectors having
$\vec\sigma\cdot \vec L^L>0$ and $-1$ if instead $\vec\sigma\cdot \vec L^L<0$. 
But if $(\vec L^L+\vec\sigma/2)^2=(L+1/2)(L+3/2)$, then 
$\vec\sigma\cdot \vec L^L=L>0$, while if 
 $(\vec L^L+\vec\sigma/2)^2=(L-1/2)(L+1/2)$, $\vec\sigma\cdot \vec L^L=
-(L+1)<0$. $\Gamma$ is $+1$ on former states and $-1$ on latter states.
Thus
\be sign\,(\vec\sigma\cdot \vec L^L)=\Gamma\ ,
\label{gwfxxx}
\ee
and similarly 
\be sign\,(\vec\sigma\cdot \vec L^R)=-\Gamma'\ .
\label{gwfxxxi}
\ee

We omit the calculation of the spectrum of $D$ as it has been done before,
see \cite{klimcik}\cite{bal2} and references there. We emphasize that
this spectrum agrees completely with the spectrum of the continuum
Dirac operator, except at the $j=(2L+1/2)$ level.

\subsection{The Fuzzy Gauged Dirac Operator (No Instanton Fields)}

We adopt the convention that gauge fields are built from operators on
$Mat(2L+1)$ which act by left multiplication. For $U(k)$ gauge theory,
we start from $Mat(2L+1)\otimes{\mathbb C}^k$. 
The fuzzy gauge fields $A_i^L$
are $k\times k$ matrices $[(A^L_i)_{mn}]$ where each entry is the operator
of left-multiplication by $(A_i)_{mn}\in\;Mat(2L+1)$ on $Mat(2L+1)$. $A^L_i$
thus acts on $\xi=(\xi_1,\ldots,\xi_k), \ \xi_i\in\;Mat(2L+1)$ according to
\be
(A_i^L\xi)_m=(A_i)_{mn}\xi_n\ .
\label{gwfxxxii}
\ee
The gauge-covariant derivative is then
\be
\nabla_i(A^L)={\cal L}_i+A^L_i=L_i^L-L_i^R+A_i^L\ .
\label{gwfxxxiv}
\ee
Note how only the left angular momentum is augmented by a gauge field.

The hermiticity condition on $A^L_i$ is 
\be
(A^L_i)^*=A^L_i\ ,
\label{gwfxxxv}
\ee
where
\be
((A^L_i)^*\xi)_m=(A^*_i)_{nm}\xi_n\ ,
\label{gwfxxxvi}
\ee
$(A^*_i)_{nm}$ being hermitean conjugate of $(A_i)_{nm}$.
The corresponding field strength $F_{ij}$ is defined by
\be
[(L+A)^L_i,(L+A)^L_j]=i\epsilon_{ijk}(L+A)^L_k+iF_{ij}\ .
\label{gwfxxxvib}
\ee

There is a further point to attend to. We need a gauge-invariant condition
which in the continuum limit eliminates the component of $A_i$ normal
to $S^2$. There are different such conditions, the following simple one
being due to \cite{nair}:
\be
(L^L_i+A^L_i)^2=(L^L_i)^2=L(L+1)\ .
\label{gwfxxxvii}
\ee
This is gauge invariant. For large $L$ it gives
\be
[x^L_i,A^L_i]_+ + \frac{(A^L_i)^2}{L}=0\ .
\label{gwfxxxviii}
\ee
$A^L_i$ is to remain bounded as $L\to\infty$. Also $x^L_i\to\hat x_i$, the
unit normal to the sphere at $\hat x$. So in the limit, if $A^L_i\to A_i,\ 
\hat x\cdot\vec A(\hat x)=0$, as required.

The Ginsparg-Wilson system can be introduced as follows. As $\Gamma$ squares
to $\BI$, there are no zero modes for $\Gamma$ and hence for 
$\vec\sigma\cdot\vec L^L+1/2$. By continuity, for generic $\vec A^L$, its
gauged version $\vec\sigma\cdot(\vec L^L+\vec A^L)+1/2$ also has no zero
modes. Hence we can set 
\be
\Gamma(A^L)=\frac{\vec\sigma\cdot(\vec L^L+\vec A^L)+1/2}{
|\vec\sigma\cdot(\vec L^L+\vec A^L)+1/2|}\ ,\quad
\Gamma(A^L)^*=\Gamma(A^L)\ ,\quad \Gamma(A^L)^2=\BI\ .
\label{gwfxxxix}
\ee
It is the gauged involution which reduces to $\Gamma=\Gamma(0)$ for
zero $\vec A^L$.

As for the second involution $\Gamma'(A^L)$, we can set
\be
\Gamma'(A^L)=\Gamma'(0)\equiv\Gamma'
\label{gwfxxxixa}
\ee

On following (\ref{gwfvi}-\ref{gwfix}), these idempotents generate the
Ginsparg-Wilson algebra with operators $\Gamma_\lambda(A^L)$, where
$\Gamma_\lambda(0)=\Gamma_\lambda$. 

The operators $\vec L^{L,R}$ do not individually have continuum 
limits as their
squares $L(L+1)$ diverge as $L\to\infty$. In contrast $\vec{\cal L}$ and
$\vec A^L$ do have continuum limits. This was remarked earlier on for the
latter, while $\vec{\cal L}$ just becomes orbital angular momentum. 

To see more precisely how  $D(A^L)$, the Dirac operator for gauge field 
$A^L$, ($D(0)$
being $D$ of (\ref{gwfxxviii})), and  $\Gamma_2(A^L)$, behave in the 
continuum limit, we note that from (\ref{gwfxxxvib}),(\ref{gwfxxxvii}) 
\be
\big(\vec\sigma\cdot(\vec L^L+\vec A^L)+\frac{1}{2}\big)^2=
(L+\frac{1}{2})^2 - \frac{1}{2}\epsilon_{ijk}\sigma_iF_{ij}\ ,
\label{gwfxla}
\ee
and therefore we have the expansions 
\be
\frac{1}{|\vec\sigma\cdot(\vec L^L+\vec A^L)+\frac{1}{2}|}
=\frac{2}{\sqrt\pi}\int_0^\infty ds\,e^{-s^2
(\vec\sigma\cdot(\vec L^L+\vec A^L)+\frac{1}{2})^2}=\frac{1}{L+\frac{1}{2}}
+\frac{1}{4(L+\frac{1}{2})^3}\epsilon_{ijk}\sigma_iF_{jk}+...,
\label{gwfxlb}
\ee
\beqa
D(A^L)&=&(2L+1)\Gamma_1(A^L)=
\vec\sigma\cdot(\vec L^L-\vec L^R+\vec A^L)+1 \;+
\frac{\vec\sigma\cdot(\vec L^L+\vec A^L)+\frac{1}{2}}{4(L+\frac{1}{2})^2}
\epsilon_{ijk}\sigma_kF_{ij}+..   \nn\\
\Gamma_2(A^L)&=&\frac{\vec\sigma\cdot(\vec L^L+\vec A^L)+\frac{1}{2}}
{2(L+\frac{1}{2} )}-
\frac{-\vec\sigma\cdot\vec L^R+\frac{1}{2}}{2(L+\frac{1}{2})}
\;+\;\frac{\vec\sigma\cdot(\vec L^L+\vec A^L)+\frac{1}{2}}{8(L+\frac{1}{2})^3}
\epsilon_{ijk}\sigma_kF_{ij}+...\ .\nn\\
\label{gwfxlc}
\eeqa
So in the continuum limit,
 $D(A^L)\to\vec\sigma\cdot(\vec{\cal L}+\vec A)+1\,$,
and $\Gamma_2(A)\to\vec\sigma\cdot\hat x$, exactly as we want.

It is remarkable that even in the presence of gauge field, there is the 
operator
\be
\Gamma_0(\vec A^L)=\frac{1}{2}[\Gamma(\vec A^L),\Gamma'(\vec A^L)]_+
\label{gwfxl}
\ee
which is in the centre of $\cal A$. It assumes the role of $\vec J^2$ in the
presence of $\vec A^L$. In the continuum limit, it has the following 
meaning.
With $D(A^L)$ denoting the Dirac operator for gauge field $A^L$, ($D(0)$
being $D$ of (\ref{gwfxxviii})), $sign\,(D(A^L))$ and $\Gamma_2(A^L)$ 
generate
a Clifford algebra in that limit and the Hilbert space splits into a direct sum
of subspaces, each carrying its IRR. $\Gamma_0(A^L)$ is a label for these 
subspaces.

\section{The Basic Instanton Coupling}
\setcounter{equation}{0}

The instanton sectors on $S^2$ correspond to $U(1)$ bundles thereon.
The connection on these bundles is not unique. Those with maximum
symmetry have a particular simplicity and are therefore important 
for analysis.

In a similar way, on $S^2_F$, there are projective modules which in the
algebraic approach substitute for sections of bundles 
\cite{connes}\cite{bal1}\cite{landi}\cite{baez}. 
There are particular connections on these modules
with maximum symmetry and simplicity. In this section we build the 
Ginsparg-Wilson system for such connections. The Dirac operator then is
also simple. It has zero modes which are responsible for the axial anomaly.
Their presence will also be shown by simple reasoning.

To build the projective module for Chern number $2T$, $T>0$, introduce 
${\mathbb C}^{2T+1}$ carrying the angular momentum $T$ representation of 
$SU(2)$. Let $T_\alpha,\ \alpha=1,2,3$ be the angular momentum operators in
this representation with standard commutation relations. Let
$Mat(2L+1)^{2T+1}\equiv Mat(2L+1)\otimes{\mathbb C}^{2T+1}$. We let $P^{(L+T)}$
be the projector coupling left angular momentum operators $\vec L^L$ with
$\vec T$ to produce maximum angular momentum $L+T$. Then the projective module 
$P^{(L+T)}Mat(2L+1)^{2T+1}$ is the fuzzy analogue of sections of $U(1)$ 
bundles on $S^2$ with Chern number $2T>0$ \cite{baez}. If instead we 
couple $\vec L^L$ and $\vec T$ to produce the least angular momentum 
$(L-T)$ using the projector
$P^{(L-T)}$, $P^{(L-T)}Mat(2L+1)^{2T+1}$ corresponds to Chern number $-2T$ (we
assume that $L\ge T$).
 
We go about as follows to set up the Ginsparg-Wilson system. For $\Gamma$
we now choose
\be
\Gamma^\pm=\frac{\vec\sigma\cdot(\vec L^L+\vec T)+1/2}{L\pm T+1/2}
\label{gwfxli} 
\ee
 The domain of $\Gamma^\pm$ is
$P^{(L\pm T)}Mat(2L+1)^{2T+1}\otimes{\mathbb C}^2$ with $\sigma$ acting on 
${\mathbb C}^2$. On this module $(\vec L^L+\vec T)^2=(L\pm T)(L\pm T+1)$ 
and $(\Gamma^\pm)^2=\BI$. 

As for $\Gamma'$, we choose it to be the same as in eq.(\ref{gwfxxvii}).

$\Gamma^\pm$ and $\Gamma'$ generate the new Ginsparg-Wilson system. 
The operators
$\Gamma_\lambda$ are defined as before as also the new Dirac operator 
$D^{(L\pm T)} = \frac{2}{a}\Gamma_1$. For $T>0$ it is convenient to choose
\be
a=\frac{1}{\sqrt{(L+\half)(L\pm T+\half)}}\ .
\label{gwfxlia}
\ee

\subsection{Mixing of Spin and Isospin}

The total angular momentum $\vec J$ which commutes with $P^{(L\pm T)}$
and hence acts on\\ $P^{(L\pm T)}Mat(2L+1)\otimes{\mathbb C}^2$ is not
$\vec L^L-\vec L^R+\vec\sigma/2$, but $\vec L^L+\vec T-\vec L^R+\vec\sigma/2$.
The addition of $\vec T$ here is the algebraic analogue
of the `mixing of spin and isospin' \cite{jackiw}. Such a term is essential
in $\vec J$ since $\vec L^L-\vec L^R+\vec\sigma/2$, not commuting with
$P^{(L\pm T)}$, would not preserve the modules. 
It is interesting that a mixing 
of `spin and isospin' occurs already in our finite-dimensional matrix model
and does not need noncompact spatial slices and spontaneous symmetry
breaking.

\subsection{ The Spectrum of the Dirac operator}

The spectrum of $\Gamma_1$ and $D^{(L\pm T)}$ can be derived
simply by angular momentum addition, confirming the results of section 2. 

On the   $P^{(L\pm T)}Mat(2L+1)^{2T+1}$ modules,
$(\vec L^L+\vec T)^2$ has the fixed values $(L\pm T)(L\pm T+1)$, and 
\beqa
(\Gamma_1)^2&=&\frac{1}{(2(L\pm T)+1)(2L+1)}\big((\vec L^L+\vec T-\vec L^R+
\frac{1}{2}\vec\sigma)^2+\frac{1}{4}-T^2\big)\ ,
\label{ai}
\\
\Gamma^\pm&=&\frac{(\vec L^L+\vec T+\frac{1}{2}\vec\sigma)^2-(L\pm T)(L\pm T+1)
-\frac{1}{4}}{(L\pm T)+\frac{1}{2}}\ ,
\label{aii}\\
\Gamma'&=&\frac{(-\vec L^R+\frac{1}{2}\vec\sigma)^2-L(L+1)-\frac{1}{4}}{L+
\frac{1}{2}}\ .
\label{aiii}
\eeqa
Comparing (\ref{ai}) with (\ref{gwfx}) we see that the `total angular 
momentum' $(\vec J)^2=(\vec L^L+\vec T-\vec L^R+\frac{1}{2}\vec\sigma)^2$ 
is linearly related to $\Gamma_0=\frac{1}{2}[\Gamma^\pm,\Gamma']_+$. 
The eigenvalues $(\gamma_1)^2$ of $(\Gamma_1)^2$ 
are determined by those of  $(\vec J)^2$, call them $j(j+1)$.

For $j=j_{max}=L\pm T+L+
\frac{1}{2}$ we have $(\Gamma_1)^2=1$, so this is 
 $V_{+1}$, and the degeneracy is $2j_{max}+1=2(2L\pm T+1)$. The
maximum value of $j$ can be achieved only if
\be
(\vec L^L+\vec T+\frac{1}{2}\vec\sigma)^2=(L\pm T+\frac{1}{2})
(L\pm T+\frac{3}{2})\ ,\quad (-\vec L^R+
\frac{1}{2}\vec\sigma)^2=(L+\frac{1}{2})(L+\frac{3}{2})\ .
\label{aiiia}
\ee 
Replacing these values in (\ref{aii},\ref{aiii}) we see that 
on $V_{+1}$ we have $\gamma_1=1$, and $\Gamma_2=0$.

The case $T=0$ has been treated before \cite{klimcik}\cite{baez}\cite{bal2}.
So we here assume that $T>0$. In that case,
for either module $j_{min}=T-\frac{1}{2}$, which gives an eigenvalue
$(\gamma_1)^2=0$ with degeneracy $2T$; we are in $V_{-1}$, the space of the 
zero modes. To realize this minimum value of $j$ we must have
\be
(\vec L^L+\vec T+\frac{1}{2}\vec\sigma)^2=(L\pm T\mp\frac{1}{2})
(L\pm T\mp\frac{1}{2}+1)\ ,\quad (-\vec L^R+
\frac{1}{2}\vec\sigma)^2=(L\pm\frac{1}{2})(L\pm\frac{1}{2}+1)\ .
\label{aiiib}
\ee
Replacing these values in (\ref{aii}, \ref{aiii}) we find that on the 
corresponding eigenstates $\Gamma_2=\mp 1$: they are all either chiral left or 
chiral right. These are the results needed by continuum index theory and axial
anomaly.

For $j_{min}<j<j_{max}$, that is on $V$, we have $0<(\gamma_1)^2<1$, and by 
(\ref{gwfxii}), $\Gamma_2\ne 0$. Since 
$[\Gamma_1,\Gamma_2]_+=0$, to each state $\psi$ such that $\Gamma_1\psi=
\gamma_1\psi$ corresponds a state $\psi'=\Gamma_2\psi$ such that 
$\Gamma_1\psi'=-\gamma_1\psi'$.

For any value of $j$ we can write $j=n+T-\frac{1}{2}$ with $n=0,1,...,
2L+1$ when the projector is $P^{(L+T)}$, and  $n=0,1,...,2(L-T)+1$ 
when the projector is $P^{(L-T)}$, while correspondingly,
\be
(\gamma_1)^2=\frac{n(n+2T)}{(2(L\pm T)+1)(2L+1)}\ .
\label{aiv}
\ee
With the choice (\ref{gwfxlia}) for $a$ this gives for the squared
Dirac operator the eigenvalues $\rho^2=n(n+2T)$. This spectrum agrees 
{\it exactly} with what 
one finds in the continuum \cite{bassetto}, except at the top value of $n$. 
Such a result is true also for $T=0$ \cite{bal2}\cite{baez}. For the top value
of $n$, $\Gamma_2=0$, and we get only the eigenvalue $\gamma_1=1$, whereas in 
the continuum, $\Gamma_2\ne 0$ and both eigenvalues $\gamma_1=\pm 1$ occur. 
This result \cite{bal2}\cite{baez}, valid also for $T=0$,
 has been known for a long time. 

Finally, we can check that summing the degeneracies of the eigenvalues we have 
found, we get exactly the dimension of the corresponding module. In fact:
\beqa
2T+2\sum_{n=1}^{2L}\big(2(n+T-\half)+1\big)+2(2L+T+1)&=&2(2L+1)(2(L+T)+1)\ ,
\nn\\
2T+2\sum_{n=1}^{2(L-T)}\big(2(n+T-\half)+1\big)+2(2L-T+1)&=&2(2L+1)(2(L-T)+1)
\ .\nn\\
\label{avi}
\eeqa

We show below that the axial anomaly on $S^2_F$ is stable
against perturbations compatible with the chiral properties of the
Dirac operator, and is hence a `topological' invariant.

\section{ Gauging the Dirac Operator in Instanton Sectors}
\setcounter{equation}{0}

The operator $\vec{\cal L}+\vec T$ commutes with $P^{(L\pm T)}$
and hence preserves the projective modules. It is important to
preserve this feature on gauging as well. So the gauge field $\vec
A^L$ is taken to be a function of $\vec L^L+\vec T$ (which remains
bounded as $L\to\infty$). For $L\to\infty$, it becomes a function of
$x$. The limiting transversality of $\vec T+\vec A^L$ can be
guaranteed by imposing the condition 
\be 
(\vec L^L+\vec T+\vec A^L)^2=(\vec L^L+\vec T)^2=(L\pm T)(L\pm T+1)\ ,
\label{gwfil}
\ee   
which generalizes (\ref{gwfxxxvii}).

We can now construct the Ginsparg-Wilson system using
\be
\Gamma(A^L)=\frac{\sigma\cdot(\vec L^L+\vec T+\vec A^L)+1/2}
{|\sigma\cdot(\vec L^L+\vec T+\vec A^L)+1/2|}
\label{gwfl}
\ee
 and the $\Gamma'$ of (\ref{gwfxxvii}), $\Gamma(0)$ being $\Gamma$ of 
(\ref{gwfxli}). $\sigma\cdot(\vec L^L+\vec T)+1/2$ has no zero modes,
 and therefore (\ref{gwfl}) is well-defined for generic $\vec A^L$. We
 can now use section 2 to construct the Dirac theory. 

We have a continous number of Ginsparg-Wilson algebras labeled by
$\vec A^L$. For each, (\ref{gwfxxi}) holds:
\be
\tr\Gamma_2(A^L)=n(A^L)\ .
\label{gwfli}
\ee
Here as $n(A^L)\in\mathbb Z$, it is in fact a constant by
continuity. The index of the Dirac operator and the axial anomaly (\ref{gwfli})
are thus independent of $\vec A^L$ as previously indicated.

The  expansions (\ref{gwfxla}-\ref{gwfxlc}) are easily extended
to the instanton sectors, and imply
the continuum limit of $D^{(L\pm T)}(\vec A^L)$ and chirality
$\Gamma_2(\vec A^L)$
\beqa 
D^{(L\pm T)}(\vec A^L)\ &\to &\ \vec\sigma\cdot(\vec{\cal L}+\vec T+\vec
A)+ 1\ ,\nn\\
 \Gamma_2(A^L)\ &\to& \ \vec\sigma\cdot\hat x\ .
\label{gwflii}
\eeqa
 Chirality is thus independent of the gauge field  in the limiting
 case, but not otherwise.

\section{Remarks}.
\setcounter{equation}{0}

The Ginsparg-Wilson system developed above can be generalized to any
number of products of $S^2_F$. For example consider  $S^2_F \otimes
S^2_F$. Its algebra is $Mat(2L+1)\otimes_{\mathbb C}Mat(2L'+1)$, where
$L$ and $L'$ can differ. There is a Ginsparg-Wilson system for each
factor with its $\Gamma$ and $\Gamma'$. Denote them (with or without
instantons and/or gauge fields) by $\Gamma(1),\Gamma'(1),\Gamma(2)$
and $\Gamma'(2)$. The $\Gamma$ and $\Gamma'$ for $S^2_F \otimes
S^2_F$ are
\beqa 
\Gamma=\frac{\Gamma(1)+\Gamma_3(1)\Gamma(2)}{|\sqrt{1+\Gamma_3(1)^2}|}\quad &,&\quad
\Gamma'=\frac{\Gamma'(1)+\Gamma_3(1)\Gamma'(2)}{|\sqrt{1+\Gamma_3(2)^2}|}\ ,\nn\\
\Gamma_3(1)=\frac{1}{2i}[\Gamma(1),\Gamma'(1)]\quad &,&
\Gamma_3(2)=\frac{1}{2i}[\Gamma(2),\Gamma'(2)]\ .
\label{gwfliii}
\eeqa
They square to unity since $[\Gamma(j),\Gamma_3(j)]_+=0$
Since the denominators in (\ref{gwfliii}) commute with the operators in
the numerators, there is no ordering problem in these equations.

Generalizations of the present investigation to fuzzy spaces such as
${\mathbb C}P^N_F$ await future work.

We have already treated integrated $U(1)_A$-anomaly. Its local form
has not been treated in the present approach, see however
\cite{peter}\cite{bg}\cite{aoki}. As for gauge anomalies, the central and
familiar problem is that noncommutative algebras allow gauging only
by the particular groups $U(N)$, and that too by their particular
representations \cite{peter}. This is so in a naive approach. There
are clever methods to overcome this problem on the Moyal planes \cite{wess}
using the Seiberg-Witten map \cite{seiberg}, but they too have failed
us for the fuzzy spaces. Thus gauge anomalies can be studied for fuzzy
spaces only in a very limited manner, but even this is yet to be
done. More elaborate issues like anomaly cancellation in a fuzzy
version of the standard model have to wait till the above mentioned
problem is solved.

\centerline{Acknowledgments}

This work was supported by DOE and NSF under contract numbers 
DE-FG02-85ER40231 and INT-9908763, and
by INFN and MIUR, Italy. We have greatly benefited by our network of
collaborators at Bratislava, Chennai, UC Davis, Dublin, Mexico and Syracuse.

\bigskip\bigskip
\bibliographystyle{unsrt}

\end{document}